\documentclass[iop,twocolumn]{aastex631}
\usepackage{amsmath}

\usepackage{gensymb}
\usepackage{hyperref}
\usepackage{graphicx,latexsym}

\newcommand\xone{x_1}
\newcommand\xtwo{x_2}
\newcommand\xthree{x_3}

\newcommand\Msun{\; {\rm M}_{\odot}}

\newcommand\kms{\; {\rm km/s}}
\newcommand\pc{\;{\rm pc}}

\newcommand\kpc{\;{\rm kpc}}

\newcommand\Gyr{\;{\rm Gyr}}

\newcommand\C{{\mathcal C}}

\newcommand\simgt{\lower.5ex\hbox{$\; \buildrel > \over \sim \;$}}
\newcommand\simlt{\lower.5ex\hbox{$\; \buildrel < \over \sim \;$}}
\newcommand{\RNum}[1]{\uppercase\expandafter{\romannumeral #1\relax}}

\newcommand\BS{$A_2$}
\newcommand\tbar{$\tau_\mathrm{bar}$}
\newcommand\PS{$\Omega_p$}

\newcommand\massratio{$M_{\mathrm{per}}/M_{\mathrm{gal}}$}
\newcommand\massratioeq[1]{$M_{\mathrm{per}}/M_{\mathrm{gal}}=#1$}
\newcommand\agama{{\sc agama}}
\newcommand\revision[1]{#1}

\def\spose#1{\hbox to 0pt{#1\hss}}
\def\dt{\spose{\raise 1.0ex\hbox{\hskip2pt$\mathchar"201$}}}

\defcitealias{Zheng2025}{Paper I}

\shorttitle{growth timescale of tidal bars}
\shortauthors{Zheng et al.}

\begin{document}

\title{Comparison of bar formation mechanisms. II. does a tidally \revision{induced} bar grow faster than an internally developed bar?}

\author[0000-0001-7707-5930]{Yirui Zheng}
% \correspondingauthor{Yirui Zheng}
% \email{yiruizheng@sjtu.edu.cn}
\affiliation{Department of Astronomy, School of Physics and Astronomy, Shanghai Jiao Tong University, 800 Dongchuan Road, Shanghai 200240, P.R. China}
% \affiliation{Key Laboratory for Particle Astrophysics and Cosmology (MOE) / Shanghai Key Laboratory for Particle Physics and Cosmology, Shanghai 200240, P.R. China}
% \affiliation{National Key Laboratory of Dark Matter Physics / Shanghai Key Laboratory for Particle Physics and Cosmology, Shanghai 200240, P.R. China}
\affiliation{State Key Laboratory of Dark Matter Physics, School of Physics and Astronomy, Shanghai Jiao Tong University, Shanghai 200240, People's Republic of China}

\author[0000-0001-5604-1643]{Juntai Shen}
\correspondingauthor{Juntai Shen}
\email{jtshen@sjtu.edu.cn}
\affiliation{Department of Astronomy, School of Physics and Astronomy, Shanghai Jiao Tong University, 800 Dongchuan Road, Shanghai 200240, P.R. China}
% \affiliation{Key Laboratory for Particle Astrophysics and Cosmology (MOE) / Shanghai Key Laboratory for Particle Physics and Cosmology, Shanghai 200240, P.R. China}
% \affiliation{National Key Laboratory of Dark Matter Physics / Shanghai Key Laboratory for Particle Physics and Cosmology, Shanghai 200240, P.R. China}
\affiliation{State Key Laboratory of Dark Matter Physics, School of Physics and Astronomy, Shanghai Jiao Tong University, Shanghai 200240, People's Republic of China}

\author[0000-0002-1378-8082]{Xufen Wu}
\affiliation{Department of Astronomy,
University of Science and Technology of China, Hefei 230026, P.R. China}
\affiliation{School of Astronomy and Space Science, University of Science and Technology of China, Hefei 230026, P.R. China}

\author[0000-0001-8962-663X]{Bin-Hui Chen}
\affiliation{Tsung-Dao Lee Institute, Shanghai Jiao Tong University, Shanghai 200240, People's Republic of China}
\affiliation{Department of Astronomy, School of Physics and Astronomy, Shanghai Jiao Tong University, 800 Dongchuan Road, Shanghai 200240, P.R. China}
% \affiliation{Key Laboratory for Particle Astrophysics and Cosmology (MOE) / Shanghai Key Laboratory for Particle Physics and Cosmology, Shanghai 200240, P.R. China}
% \affiliation{National Key Laboratory of Dark Matter Physics / Shanghai Key Laboratory for Particle Physics and Cosmology, Shanghai 200240, P.R. China}
\affiliation{State Key Laboratory of Dark Matter Physics, School of Physics and Astronomy, Shanghai Jiao Tong University, Shanghai 200240, People's Republic of China}

%===============================================================================

\begin{abstract}

Bar structures can form internally due to the instability of their host galaxies or externally due to perturbations from other galaxies. 
We systematically quantify the growth timescales (\tbar)  of bars formed through these two mechanisms with a series of controlled $N$-body simulations.
% We employ exponential fitting to quantify the growth timescale \tbar\ of the bar strength following the theoretical prediction of the swing amplification mechanism.
In galaxies susceptible to bar instability, tidally \revision{induced} bars display \tbar\ values comparable to those of internally developed bars within the same disk.
% When tidal perturbations either promote or delay bar formation, the onset of the formation is accordingly advanced or postponed, but the growth rate of the bar structure remains largely unchanged. 
Tidal perturbations promote(delay) bar formation by advancing(postponing) its onset,
but the growth rate of the bar structure remains largely unchanged.
In these interaction scenarios, the bar formation is still driven primarily by the galaxy's internal nature, which remains unaffected by tidal perturbations.
As the external perturbation wave reaches the galaxy's center, it evokes a ``seed bar'' that is then swing amplified. In this scenario, the onset of bar formation is advanced. Conversely, bar formation may be delayed if the external perturbation wave is out of phase with the preexisting spontaneously developed seed bar, which causes destructive interference and limits the bar growth.
In the hot disk model that resists bar formation in isolation, the \tbar\ of the tidally \revision{forced} bar correlates with the strength of the perturbation. 
The bar growth in this model deviates from an exponential profile and is better described by a linear function. 
% The strictly-speaking ``tidally \revision{forced} bars'' 
% \revision{The tidally \revision{forced} bars in the hot disk model}
% may not adhere to the swing amplification mechanism that predicts an exponential bar growth. Their preference for linear growth contrasts with bars formed in galaxies inherently susceptible to bar instability.
\revision{The varied \tbar\ and the preference for linear growth contrast with bars formed in galaxies inherently susceptible to bar instability.
These tidally \revision{forced} bars 
may not adhere to the swing amplification mechanism that predicts an exponential bar growth.
}

\end{abstract}

\keywords{%
  galaxies: kinematics and dynamics ---
  galaxies: structures
}

%-----------
%-- Sect. 1
%-----------

\section{Introduction}
\label{sec:intro}

% - different formation mechanisms

% - Interactions can either promote/delay bar formation

% - swing amplifier feedback loop

% - exponential growth and growth timescale

% Bars, including both strong and weak bars, have been found in over 60\% of disk galaxies in the local universe (de Vaucouleurs et al. 1991; Buta et al. 2010, 2015; Ann et al. 2015). The bar fraction has been widely known to depend strongly on the Hubble sequence, mass, color, and bulge prominence (Sheth et al. 2008).

% Bars are a widespread feature in spiral galaxies.
Bars are a common feature in spiral galaxies. 
Locally, about half of spiral galaxies possess bars in optical bands, with this proportion rising to two-thirds in infrared bands \citep{Marinova2007, MenendezDelmestre2007, Erwin2018, Lee2019bar}. Although a smaller fraction of galaxies have bars at higher redshifts, they still represent 13\% in the range $2<z\leq 3$ \citep{LeConte2024} and appear as early as $z\sim4$ \citep{Guo2024}. 
Bars significantly influence the evolution of their host galaxies, including triggering gas inflow, affecting star formation, and contributing to the formation of pseudobulges \citep{masters2011galaxy, Li2015, Lin2017, Lin2020, Iles2022}. The Milky Way also has a bar \citep[e.g.,][]{deVaucouleurs1964, Blitz1991}, and its boxy bulge and gas dynamics are directly linked to this bar \citep[e.g.,][]{Shen2010, Li2018}. The prevalence and impact of bars on galaxies highlight the importance of studying their formation and evolution.
% in galaxy dynamics.

Galactic bars can form via internal or external mechanisms. 
Internally, bars can develop spontaneously due to the gravitational instability of the disks \citep{hohl1971numerical, ostriker1973numerical, sellwood2014secular, Lokas2019iso}. 
A widely recognized mechanism for spontaneous bar formation is the swing amplification loop, initially developed by \citet{Goldreich1965} \revision{and} \citet{Julian1966} and further elaborated by \citet{Toomre1981}. 
\citet{Binney2020} revisited this concept and provided a more accessible explanation. 
According to this model, any source of noise generates a packet of leading spiral waves, which are then amplified as they swing from leading to trailing. 
The amplified trailing waves propagate through the disk center and reemerge as a leading disturbance, which then triggers the next swing amplification loop. 
The repeated amplification process amplifies minor noise into a prominent spiral pattern and ultimately facilitates the formation of a coherent bar. 
The positive feedback loop predicts exponential growth of the spiral/bar strength.

Several studies investigated the growth rate or equivalently the formation timescale of bars formed via internal instability.
The formation timescale is found to depend on the bulge-to-disk mass fraction \citep{Kataria2018}, the Toomre $Q$ parameter \citep{Hozumi2022, Worrakitpoonpon2024}, and the disk thickness \citep{Ghosh2023}.
\citet{Fujii2018} found an exponentially decreasing relation between the bar formation timescale and \revision{increasing} disk mass fraction. \revision{The ``Fujii relation''} was then confirmed in \citet{BlandHawthorn2023}.
An important improvement in the latter work is the quantification of the exponential growth timescale associated with the swing amplification loop instead of Fujii's fixed threshold for when a bar appears ($A_{2,max}>0.2$).
\revision{\citet{BlandHawthorn2023} demonstrated that an exponential fit to the growth of \BS\ over time provides a timescale and an onset time, and also reproduces the Fujii relation independently of any assumptions about \BS.}
\revision{Several studies have suggested that the disk mass fraction plays a primary role in determining the bar formation timescale \citep{Fujii2018, BlandHawthorn2023, BlandHawthorn2024,BlandHawthorn2025}. Thus, it is important to control the disk mass fraction when investigating the effects of other parameters on bar formation.}

Besides the internal instability, bar formation can also be externally induced by various gravitational perturbations, including flyby encounters with other galaxies, collisions, mergers, or the tidal effects of galaxy clusters \citep{byrd1986tidal,gauthier2006substructure,  Lokas2016,MartinezValpuesta2017, Lokas2019tidal}.
A comprehensive investigation is desired to compare the growth and evolution of bars formed through different mechanisms. 
Studies have indicated that tidal interactions can either promote or delay bar formation \citep{Lokas2016, Moetazedian2017, Lokas2018, Pettitt2018, Zana2018}. This raises a question: if tidal forces either promote or delay the formation of bars, do they result in bars that grow at a faster or slower rate compared to those formed internally? Alternatively, do these bars grow at similar rates but onset at different times?
\citet{Moetazedian2017} found that the exponential growth rate of tidal bars is independent of the number of satellite encounters and their orbits. 
However, the rescaled mass of the primary satellite in their simulation is only \revision{$<$1}\% compared to the host halo mass of the galaxy, which limits the strength of the tidal perturbation.

% To further address this question, we employ a series of controlled $N$-body simulations described in \citet[][Paper I hereafter]{Zheng2025}, where we compare the pattern speed \PS\ of tidally \revision{forced} bars to their spontaneous counterparts.
% \revision{
Previous studies have predominantly focused on either the internal or external mechanisms of bar formation, with less attention given to comparing these two mechanisms. 
In this series of papers, we aim to offer a comprehensive comparison between bars formed through different mechanisms. 
In \citet[][Paper I hereafter]{Zheng2025}, we conducted a series of controlled $N$-body simulations to examine the pattern speed \PS\ of tidally \revision{induced} bars and compared them with their spontaneous counterparts in the same galaxy model. 
We found that bars formed via both mechanisms exhibit similar distributions in the pattern speed--bar strength ($\Omega_p-A_2$) space and display comparable ratios of corotation radius to bar length (${\cal R}={R_{\mathrm {CR}}}/{R_{\mathrm {bar}}}$). 
As for ``tidally \revision{forced} bars'' that refer specifically to those in galaxies stable against bar instability, they indeed rotate slower than internally developed ones in less stable galaxies. 
However, this difference in pattern speed is attributed to the varying internal properties of bar host galaxies, rather than the different formation mechanisms.
In this second paper, we broaden our investigation to compare the growth timescales and onset times between tidally \revision{induced} and spontaneously developed bars. Our objective is to understand how external perturbations influence the growth of bar structures. 
% }

% With the same simulation set, we investigate the growth timescale of bars formed through different mechanisms.
The structure of the paper is as follows. 
In Section \ref{sec:sims}, we provide an overview of the galaxy models and simulation details. 
Section \ref{sec:exp_growth} describes the exponential fitting method used to quantify the bar growth timescale.
In Section \ref{sec:results}, we present a comparison of the growth timescales and the onset times of bar formation between tidally \revision{induced} and internally developed bars.
The results are interpreted and discussed within the same section.
Finally, we conclude with a summary of our findings in Section \ref{sec:summary}.

%-----------
%-- Sect. 2
%-----------

\section{Simulations}
\label{sec:sims}
% \citep[][RM19 hereafter]{rodriguez-montero2019}

% In this paper, we utilize the same series of controlled $N$-body simulations described in \citetalias{Zheng2025} to investigate the bar growth timescale under different mechanisms.
% We refer the reader to \citetalias{Zheng2025} for details of the simulations. Here, we provide a brief description of the key aspects of the simulations.

In this study, we employ the same series of controlled $N$-body simulations as detailed in \citetalias{Zheng2025} to investigate the bar growth timescale under various mechanisms. For comprehensive details on the simulations, we direct the reader to \citetalias{Zheng2025}. Here, we offer a concise overview of the simulations' key aspects.

We employ the \agama\ software \citep{AGAMA2019} to generate three galaxy models. 
These models share a common density profile, differing only in the velocity distribution of their stellar disks. 
The shared density profile comprises an exponential, quasi-isothermal stellar disk and a truncated Hernquist dark matter (DM) halo. 
The stellar disk's characteristics include a mass of $M_* = 3.6 \times 10^{10} \Msun$, a scale length of $R_d = 2 \kpc$, and a scale height of $h_z = 0.4 \kpc$. 
The DM halo features a mass of $M_{\rm halo} = 3.6 \times 10^{11} \Msun$ and a scale radius of $a = 13.7 \kpc$. The combined mass of the stars and DM halo is $M_{\rm tot} = 4.0 \times 10^{11} \Msun$.

%, resulting in a disk thickness ratio of $h_z/R_d = 0.2$.
% , and $r_{\rm cut} = 114 \kpc$. 
% %fd is about 0.47, but we need to check the exact value; the old code gives a rough estimation. --> go AGAMA

\agama\ employs an action-based distribution function
(DF) to determine the velocity distribution of the stellar disk. 
We highlight the radial velocity dispersion, a critical physical quantity influencing disk stability. 
In all disk models, the radial velocity dispersion decreases exponentially with radius, with its scale length being twice that of the density profile, $R_{\sigma, R} = 2\;R_d$. 
By varying the central value of the radial velocity dispersion ($\sigma_{R,0}$) we create three disk models: cold, warm, and hot disks, characterized by $\sigma_{R,0} = 73 \kms$, $124 \kms$, and $226 \kms$, respectively.

\revision{
Following \citet{Fujii2018} and \citet{BlandHawthorn2023}, we calculate the $f_{\text{disk}}$ parameter as the ratio of the disk mass to the total galaxy mass within the radius at $R = 2.2\;R_d$ at which the rotation curve roughly peaks. $f_{\text{disk}}$ is defined by
\begin{equation}
f_{\text{disk}} =  \frac{V^2_{\text{c,disk}}(R)}{V^2_{\text{c,tot}}(R)} \Bigg\vert _ {R = 2.2\;R_d} ,
\end{equation}
with $V_{\text{c}}(R)$ denoting the circular velocity at radius $R$.
% \begin{equation}
% where $V_{\text{c,disk}}$ and $V_{\text{c,tot}}$ are the circular velocities of the disk and total mass, respectively.  
%   f_{\text{disk}} =  \left( \frac{V_{\text{c,disk}}(R)}{V_{\text{c,tot}}(R)} \right)^2 _{R = 2.2\;R_{\text{d}}}
% \end{equation}
% $R_s = 2.2 R_d$ is the traditional scale length adopted in studies of disks.
The cold, warm, and hot disks have $f_{\text{disk}}$ values of 0.47, 0.47, and 0.44, respectively, positioning them within the intermediate disk mass fraction range on the Fujii relation \citep{Fujii2018, BlandHawthorn2023}.
}

\revision{
These models have very similar $f_{\text{disk}}$ values, which is an expected outcome of using identical parameters for the stellar disk and DM halo density profiles.  
The hot disk model has a slightly lower $f_{\text{disk}}$ value due to its slightly reduced circular speed in the inner region. The rationale behind the hot disk model's lower circular speed is elaborated in Section~2.1 and Figure~1 of \citetalias{Zheng2025}.
}

\revision{With $f_{\text{disk}}$ being very similar among the models, the Toomre $Q$ parameter is important in evaluating the stability of the disks.
The $Q$ curves show a similar shape across all three disk models but hold different values.} 
The cold, warm, and hot disks exhibit minimum Toomre stability values ($Q_{\mathrm{min}}$) of 0.82, 1.34, and 2.24, respectively. 
According to the criterion for bar formation in \citet{Jang2023}, the cold disk is susceptible to bar instability, the warm disk is marginally stable, and the hot disk is stable against bar formation. 
In isolated simulations, the cold and warm disks are capable of spontaneously forming bars, with the warm disk requiring more time to reach a comparable bar strength. 
The hot disk does not develop a bar within 6\Gyr.

Utilizing these galaxy models, we perform a series of flyby interaction simulations. The perturber in these simulations is a pure Hernquist DM halo composed of live particles. 
The galaxy and perturber are placed on a hyperbolic orbit, inspired by the configuration presented by \citep[][refer to her Figure 1 for a schematic]{Lokas2018} . 
We investigate mass ratios of the perturber to the galaxy (\massratio) of 1/1, 1/3, and 1/10, resulting in strong, moderate, and weak tidal perturbations, respectively. 
For each mass ratio, we adjust the inclination angle between the stellar disk plane and the perturber's orbit plane to simulate prograde ($0^\circ$), perpendicular ($90^\circ$), and retrograde ($180^\circ$) flyby interactions. 
The closest approach is set to occur at approximately $t_{\rm peri} = 0.5$\Gyr\ for all flyby simulations.

Each simulation is evolved for 6\Gyr\ with \texttt{GADGET-4} \citep{Springel2005, Springel2021}. We populate the stellar disk with $0.5$ million particles and the galaxy DM halo with one million particles, leading to particle masses of $7.2\times10^4 \Msun$ for the disk and $3.6 \times 10^5 \Msun$ for the halo. The gravitational softening lengths are set at 23\pc\ for stars and 57\pc\ for DM particles.
\revision{We acknowledge that the mass resolution of our simulations is not exceptionally high, yet it is adequate to capture the growth of the bar.
\cite{Dubinski2009} demonstrated that the evolution of the bar exhibits convergent behavior when the number of halo particles ranges from one million to 10 million. This convergence is evident when studying bar growth, pattern speed evolution, the dark matter halo density profile, and the nonlinear analysis of orbital resonances. 
}
%-----------
%-- Sect. 3
%-----------

\section{Bar growth timescale}
\label{sec:exp_growth}

\begin{figure*}
  \centering
  \includegraphics[width=\textwidth]{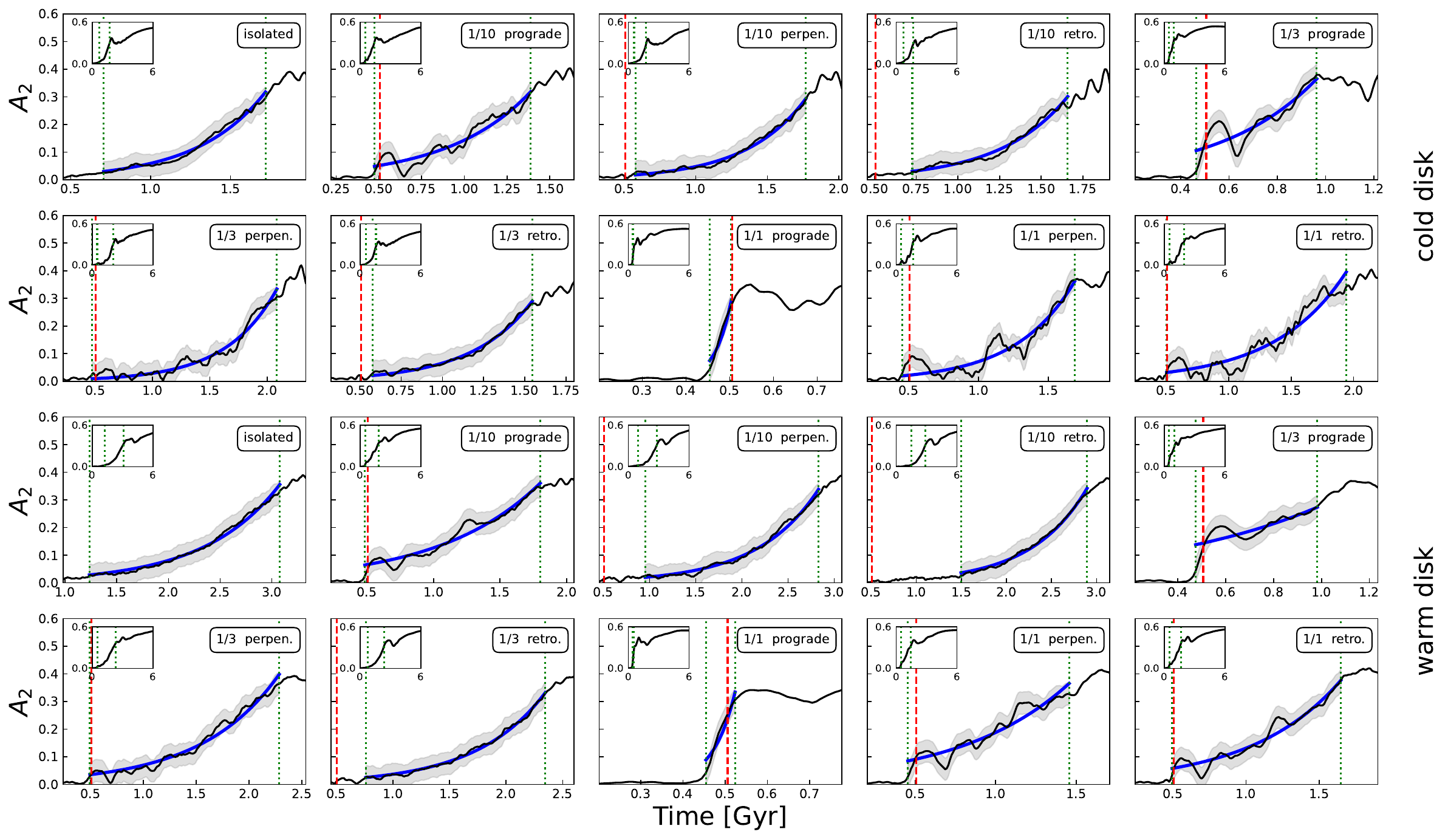}
  % \caption{test}
  \caption{Exponential fitting to the $A_2$ growth.
  The black solid lines represent the evolution of $A_2$, with the $1\sigma$ range highlighted by the gray area. The blue solid lines indicate the results of the exponential fitting. The regions selected for fitting are denoted by green vertical dotted lines, while the intended pericenter time, $t_{\rm peri}=0.5$ Gyr, is marked by red vertical dashed lines. The inset axes display the $A_2$ evolution across the entire simulation duration of 6\Gyr. The text box notes the simulation information of mass ratio (\massratioeq{1/10}, 1/3, and 1/1) and orbit type (prograde, perpendicular, and retrograde).}
\label{fig:exp_fit}
\end{figure*}

The bar strength $A_2$ is measured as the amplitude of the $m=2$ Fourier mode of the stellar disk. We first calculate the sine and cosine coefficients for the stellar disk:
\begin{subequations}
\begin{align}
  a_2 &= \frac{ \sum_{i=1}^{N} m_i \cos(2\phi_i)} {\sum_{i=1}^{N} m_i}, \\
  b_2 &= \frac{ \sum_{i=1}^{N} m_i \sin(2\phi_i)} {\sum_{i=1}^{N} m_i},
\end{align}
\end{subequations}
where $m_i$ and $\phi_i$ represent the mass and the azimuthal angle of the $i$th particle, respectively. The summation is conducted over the stellar particles within a cylindrical region where $R \leq 4\;R_d$, i.e., $R \leq 8 \kpc$ in our galaxy models. 
The bar strength $A_2$ is then calculated as
\begin{equation}
  A_2 = \sqrt{a_2^2 + b_2^2}.
\end{equation}

\revision{Following \citet{BlandHawthorn2023}, we} fit the $A_2$ evolution with an exponential function:
\begin{equation}
\label{eq:exp_fit}
A_2 (t) = 0.1  \exp((t - t_0)/\tau_{\mathrm{bar}})
\end{equation}
to obtain the growth timescale \tbar\ and the time when $A_2$ reaches 0.1 (noted as $t_0$).
\tbar\ characterizes the growth rate of the bar structure, while $t_0$ reflects the onset time of bar formation.
We note that the exact choice of $A_2$ threshold for $t_0$ should be high enough to avoid noise but low enough to capture the early phase of bar formation.
\autoref{eq:exp_fit} is equivalent to a simple exponential growth model:
\begin{equation}
A_2 = C \exp(t /\tau_{\mathrm{bar}}),
\end{equation}
where $C = 0.1 \exp(-t_0/\tau_{\mathrm{bar}})$. 
\revision{When applying to the same data set, both fitting functions yield the same value and the same standard deviation error of the growth timescale \tbar. Neither function is better than the other, but we prefer \autoref{eq:exp_fit} for fitting the $A_2$ evolution due to the slightly better physical interpretability of $t_0$ compared to $C$. Besides, the $t_0$ parameter assists in indicating the initiation time of bar formation, which is beneficial when comparing bar formation scenarios with similar \tbar\ values. This is helpful when comparing tidally induced bars with the bars formed in isolated simulations within the same disk model as discussed in \autoref{sec:results}.}
We explored other fitting functions in the \autoref{app:diff_fit}. 
These different functions yield similar \tbar\ values with differences of approximately 20\% while maintaining equivalent fitting quality.

We perform the fitting with the Levenberg--Marquardt algorithm through the \texttt{curve\_fit} function in the \texttt{scipy.optimize} package \citep{2020SciPy-NMeth} \revision{, following the same method as \citet{BlandHawthorn2023}}.
The fitting range starts when the bar strength reaches $A_2=0.025$ and ends when $A_2$ reaches $0.8\;A_{2, peak}$ with $A_{2, peak}$ being the peak $A_2$ value during the bar formation stage. 
\footnote{\revision{Only the bar formation stage before the bar buckling phase is considered for the fitting. The secular growth stage after buckling is excluded.}}
This lower limit is chosen to avoid initial stages when \BS\ cannot be well decided.
The upper limit is set to avoid the saturation of the bar strength at which the bar growth deviates from the exponential profile. 
We manually introduce a constant error\footnote{The exact value of $\sigma_{A2}$ is not crucial as long as it is reasonably chosen to avoid unrealistic impact on the fitting results.} of $\sigma_{A2}=0.04$ to the $A_2$ values to account for noise and to calculate the $1\sigma$ error of the fitting parameters.

Our fitting results are presented in \autoref{fig:exp_fit}. 
The majority of the simulations show exponential growth profiles of the bar strength with the fitting results well overlapping the $A_2$ evolution, consistent with the swing amplifier feedback loop theory.
However, several simulations show a deviation from this exponential growth, predominantly occurring in prograde interactions with massive perturbers. 
Visual examination of the stellar surface density maps reveals that this deviation is primarily due to the substantial contribution of spiral arms to the $A_2$ signal. 
\revision{We provide more details regarding our fitting method and further demonstrate its reliability in the \autoref{app:diff_fit}.}
% Despite these exceptions, the exponential fitting remains a reliable indicator of bar growth for the majority of the simulations. 
% We further validate the reliability of our fitting method by comparing the fitted $t_0$ values against the true $t_0$ in \autoref{app:diff_fit}. 
% This comparison demonstrates a strong agreement between the fitted and true $t_0$ values, with most simulations exhibiting an offset of less than 0.1\Gyr.
% \vspace{30pt}
% \hrule

% \begin{itemize}
%   \item use \( A_2 = 0.1  \exp((t - t_0)/\tau_{\mathrm{bar}}) \) to fit all simulation --> figure
  
%   \begin{itemize}
%       \item \( \tau_{\mathrm{bar}} \): the exp. growth timescale
%       \item \( t_0 \): the time when \( A_2 \) reach 0.1
%   \end{itemize}

%   \item compare fitted \( t_0 \) and the real one to illustrate the reliability of the fitting --> figure

% \end{itemize}

\section{results \& discussions}
\label{sec:results}
\subsection{Do tidal bars grow faster?}
\label{sec:tidal_growth}

% include a2ps_flyby.pdf, full page
\begin{figure*}
  \centering
  \includegraphics[width=0.85\textwidth]{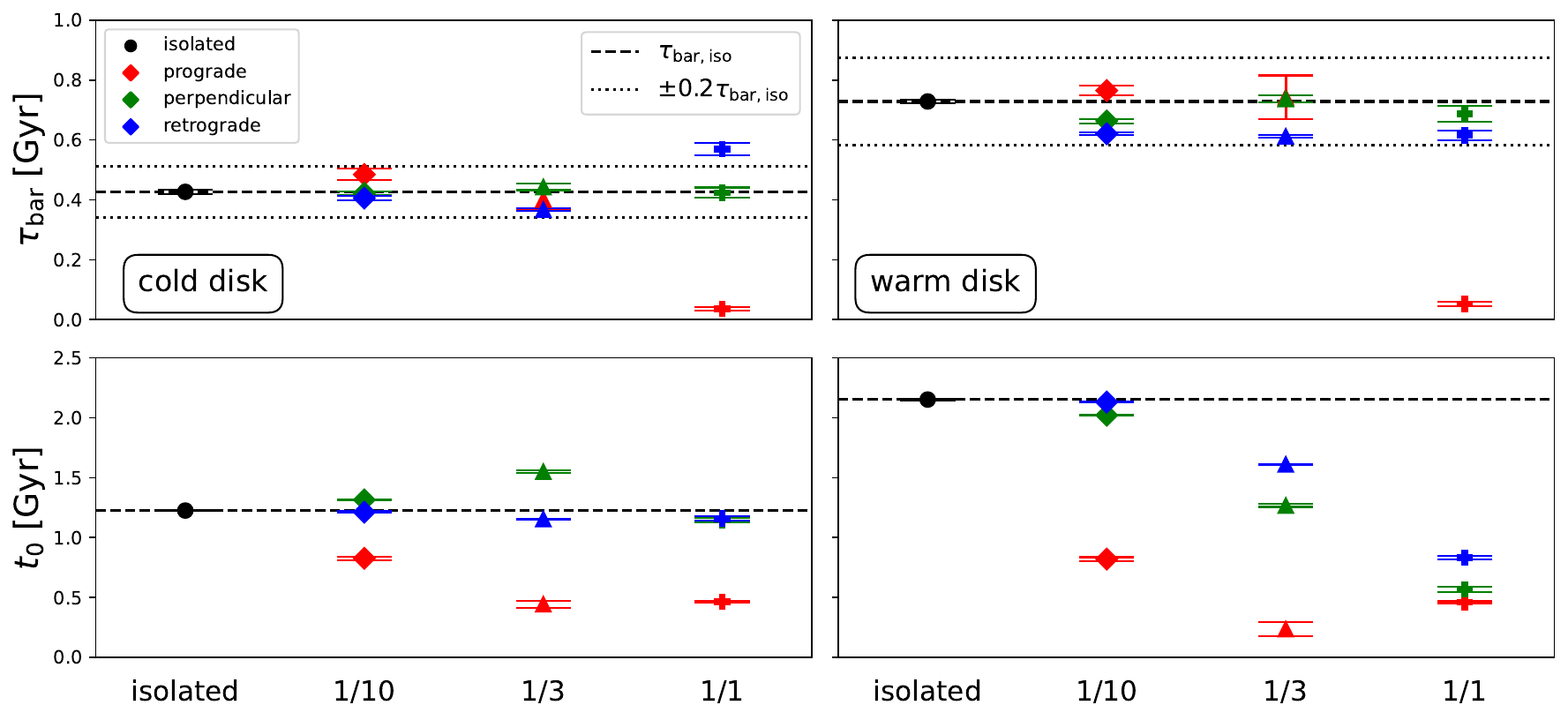}
  \caption{
    Results of the exponential fitting for the cold disk model (\textit{left column}) and the warm disk model (\textit{right column}). Each scatter point represents a simulation \revision{and the error bar shows the standard deviation on the parameter estimate.} The dashed lines indicate the results for the same disk model in isolation. \textit{Upper panels}: the growth timescale \tbar. Tidally \revision{induced} bars exhibit \tbar\ values similar to those of internally developed bars, with an offset of less than 20\%, suggesting that tidal bars and spontaneous bars in the same disk have comparable growth rates. 
    \textit{Lower panels}: the fitted $t_0$, representing the time at which $A_2$ reaches 0.1. Tidal perturbations promote bar formation by advancing the onset time, as evidenced by their smaller $t_0$ values. 
    % \textit{Bottom panels}: same as the middle panels, but for the true $t_0$.
    }
  \label{fig:fit_results}
\end{figure*}

\begin{figure*}
    \centering
    \includegraphics[width=0.85\textwidth]{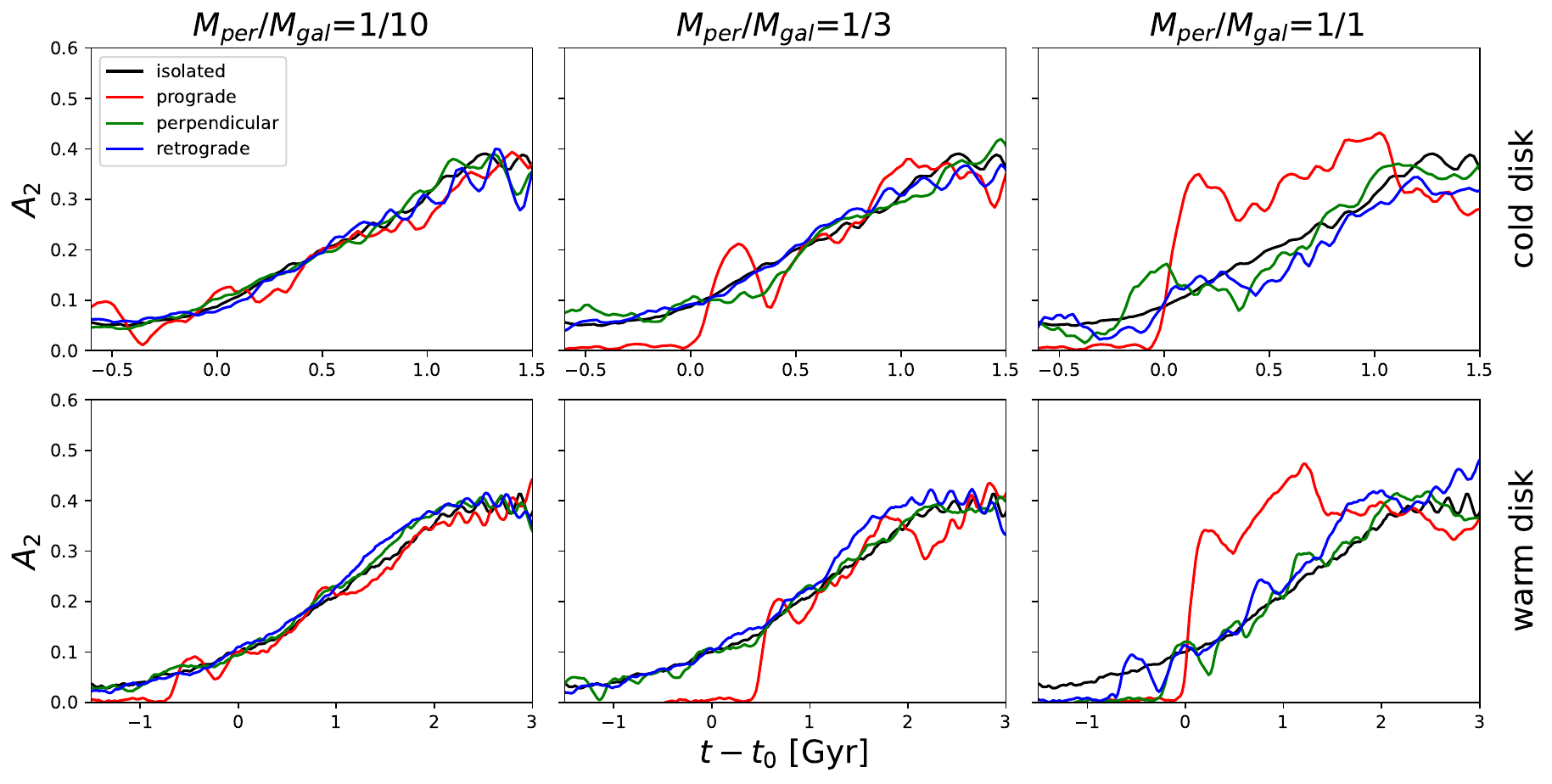}
    \caption{
      The $A_2$ evolution against \revision{$(t-t_{0})$.}
      All tidally \revision{induced} bars, except those in prograde interactions (red lines) with massive perturbers where spiral arms significantly contribute to the \BS\ signal, exhibit growth patterns very similar to the isolated case. This similarity confirms that tidal bars share comparable growth rates with their spontaneous counterparts in the same disk.
    }
    \label{fig:shifted_a2}
\end{figure*}

% \begin{figure}
%   \centering
%   \includegraphics[width=1\columnwidth]{./X_evo.pdf}
%   % \caption{test}
%   \caption{The evolution of the swing amplification parameter $X_{m=2}$ (blue) at $R=2.2\;R_d$ along with bar strength $A_2$ (black). 
%   The blue line represents the mean value of $X$, with the standard deviation indicated by the shaded region. The vertical dashed line indicates the intended time of closest approach of the perturber.
%   $X$ experiences slightly more fluctuations during the interaction but remains largely unchanged until the formation of a strong bar ($A_2>0.2$).
%   }
%   \label{fig:x_evo}
% \end{figure}  

Tidal interactions can either promote or delay bar formation.
In this section, we investigate how external perturbations influence the growth of bar structures by comparing the growth timescale \tbar\ and the onset time $t_0$ of bar formation between tidally \revision{induced} and internally developed bars within the same disk.

% \autoref{fig:fit_results} compares \tbar\ and $t_0$ for tidally \revision{forced} bars with those of internally developed bars within the same disk 
The comparison is presented in \autoref{fig:fit_results} with the \textit{left column} for the cold disk and the \textit{right column} for the warm disk.
The upper panels show that the tidally \revision{induced} bars exhibit \tbar\ comparable to that of the internally developed bars in the same disk. 
In most cases, the discrepancy is less than 20\%. 
Considering that different fitting functions can yield \tbar\ differences of approximately 20\%, our result suggests that tidal bars share similar growth rates with their spontaneous counterpart in the same disk. 
\revision{Additionally, the relatively small standard deviation of \tbar\ indicates that the fitting results are robust against noise in the $A_2$ evolution.}

The largest discrepancy in \tbar\ is observed in prograde interactions with a mass ratio of 1/1. In such cases, the perturbation is sufficiently strong to generate prominent spiral arms that significantly contribute to the $A_2$ signal. 
This situation reduces the reliability of the fitted \tbar\ as it no longer exclusively represents the growth of the bar structure. 
\revision{Bland-Hawthorn (2025, in preparation) reports bars and spiral arms may be separated by radial actions. However, it is beyond the scope of this study to disentangle the contribution of the spiral arms to the $A_2$ signal from the bar growth.} 
Therefore, we conclude that tidally \revision{induced} bars grow as fast as their spontaneous counterpart in the same disk while acknowledging the potential influence of spiral arms on the $A_2$ signal.

% In the middle panels of \autoref{fig:fit_results}, we show the fitted $t_0$, the time when the fitted $A_2$ profile reaches 0.1. 
% The tidally \revision{forced} bars have an earlier onset of their formation as indicated by their smaller $t_0$ values.
% The stronger the tidal perturbation, the earlier the onset of the bar formation.
% The only exception happens in the cold disk with the perpendicular interactions with a mass ratio of \massratio$=1/3$, where the bar formation is delayed by $\sim0.4$\Gyr.

In the lower panels of \autoref{fig:fit_results}, we present the fitted $t_0$, the time at which the fitted $A_2$ profile reaches 0.1. 
Tidally \revision{induced} bars have an earlier onset of formation, as indicated by their smaller $t_0$ values. The onset of bar formation occurs earlier with stronger tidal perturbations. 
The only exception happens in the perpendicular interaction of the cold disk with a perturber of \massratio$=1/3$, where bar formation is delayed by  $\sim0.4$\Gyr.
The underlying cause of the delay will be explored in subsection \ref{sec:onset}.

In the cold disk model (\textit{lower left panel}), the prograde interactions exhibit significantly smaller $t_0$ values compared to the isolated case, whereas the perpendicular and retrograde interactions show similar $t_0$ values. 
These results indicate that bar formation in the cold disk is predominantly driven by internal disk instability, with only severe external perturbations capable of accelerating the emergence of bars.
 
%------------------

In the warm disk model (\textit{lower right panel}), all interactions result in smaller $t_0$ values compared to the isolated case except for the weakest interactions with \massratioeq{1/10} on perpendicular and retrograde orbits. 
The warm disk is only marginally stable against bar formation in isolation. 
In its interaction simulations, bar formation is a product of the combination of both internal disk instability and external perturbations.

% In the lower panels of \autoref{fig:fit_results}, we show the true $t_0$, the time when the actual $A_2$ profile reaches 0.1.
% These results corroborate the findings from the fitted $t_0$.

\autoref{fig:fit_results} illustrates that tidal interactions promote(delay) bar formation by either advancing(postponing) the onset of this process. 
Following the onset of bar formation, tidally \revision{induced} bars grow at a rate comparable to internally developed bars within the same disk. 
% This conclusion is further reinforced by \autoref{fig:shifted_a2}, where we plot the $A_2$ evolution against $(t-t_{0, \mathrm{fit}})$. 
This conclusion is further reinforced by \autoref{fig:shifted_a2}, where we plot the $A_2$ evolution against \revision{$(t-t_{0})$. }
All tidally \revision{induced} bars exhibit similar growth patterns when referenced against the isolated case, regardless of whether the tidal perturbation accelerates or postpones bar formation. 
The only notable deviation occurs in prograde interactions with massive perturbers, where the contribution of spiral arms to $A_2$ is significant.
In such scenarios, the tidal force exerted by the perturber impacts stars in the prograde disk for a longer duration than those in the perpendicular or retrograde disk, thereby promoting the development of spiral arms. We refer the reader to \citet{Lokas2018} and \citetalias{Zheng2025} for more discussion on the effect of the inclination of the perturber's orbit.

The consistency in \tbar\ suggests that the growth of tidal bars is predominantly governed by the inherent nature of the galaxy, which remains largely unaffected by the tidal perturbation. 
Tidal perturbations do not substantially alter the swing amplification process,
although they provide stronger initial noise that triggers the swing amplification loop.
Tidally \revision{induced} bars grow as fast as their spontaneous counterpart in the same disk.

Our results suggest the fundamental mechanism driving bar growth in cold and warm disks may still be their internal bar instability even under tidal perturbations.
External perturbation mainly affect the onset of the bar formation.
Bars formed under tidal interactions in these disks are not purely tidally \revision{induced} but rather a result of both internal instability and external perturbation.
For convenience, we continue to use the term ``tidally \revision{induced} bars'' loosely throughout this paper.

These results concerning the \textit{growth timescale comparison} align with the findings in \citetalias{Zheng2025}, where we reached the same conclusion through a \textit{pattern speed comparison}. 
\citetalias{Zheng2025} showed that tidally \revision{induced} bars rotate at speeds comparable to those of internally developed bars within the same disk when bar strength and/or length are taken into account. 
Both \citetalias{Zheng2025} and this work suggest that tidal perturbations influence the bar's evolution by advancing or delaying it to a different stage, but they do not modify the bar's intrinsic properties or underlying formation mechanism.

Note that \tbar\ is not precisely identical between tidally \revision{induced} bars and internally developed ones. 
We propose two potential factors for this minor discrepancy. 
One is the transient spiral arms that contribute to the $A_2$ signal with the extreme cases being those of the prograde interactions with a mass ratio of 1/1.
% {\textcolor{red}{The other factor is the effect of the DM halo spin that is not caught by the swing amplification parameter $X$.}}
The other factor is the effect of the DM halo spin.
Although the primary galaxy's halo is initially set up without net angular momentum, the tidal force from the perturber spins it up. 
Isolated simulations have demonstrated the significance of inner halo angular momentum in bar formation: bars grow more rapidly in prograde halos while more slowly in counterrotating halos \revision{\citep[see e.g.][]{Collier2021,Kataria2022,Joshi2024}}. 
Observations also suggest a relationship between bar structures, DM halo mass and spin, and galaxy mass and spin \citep{Romeo2023}. 
The counterrotating halo spin may explain why the retrograde 1/1 interaction of the cold disk has a larger \tbar.

Although likely a secondary factor, the spun-up halo does contribute to bar formation. A detailed examination of the halo spin effect in interactions will be addressed in subsequent papers of this series.

\subsection{How perturbation affects the onset of the bar formation}
\label{sec:onset}

\begin{figure*}
  \centering
  \includegraphics[width=0.95\textwidth]{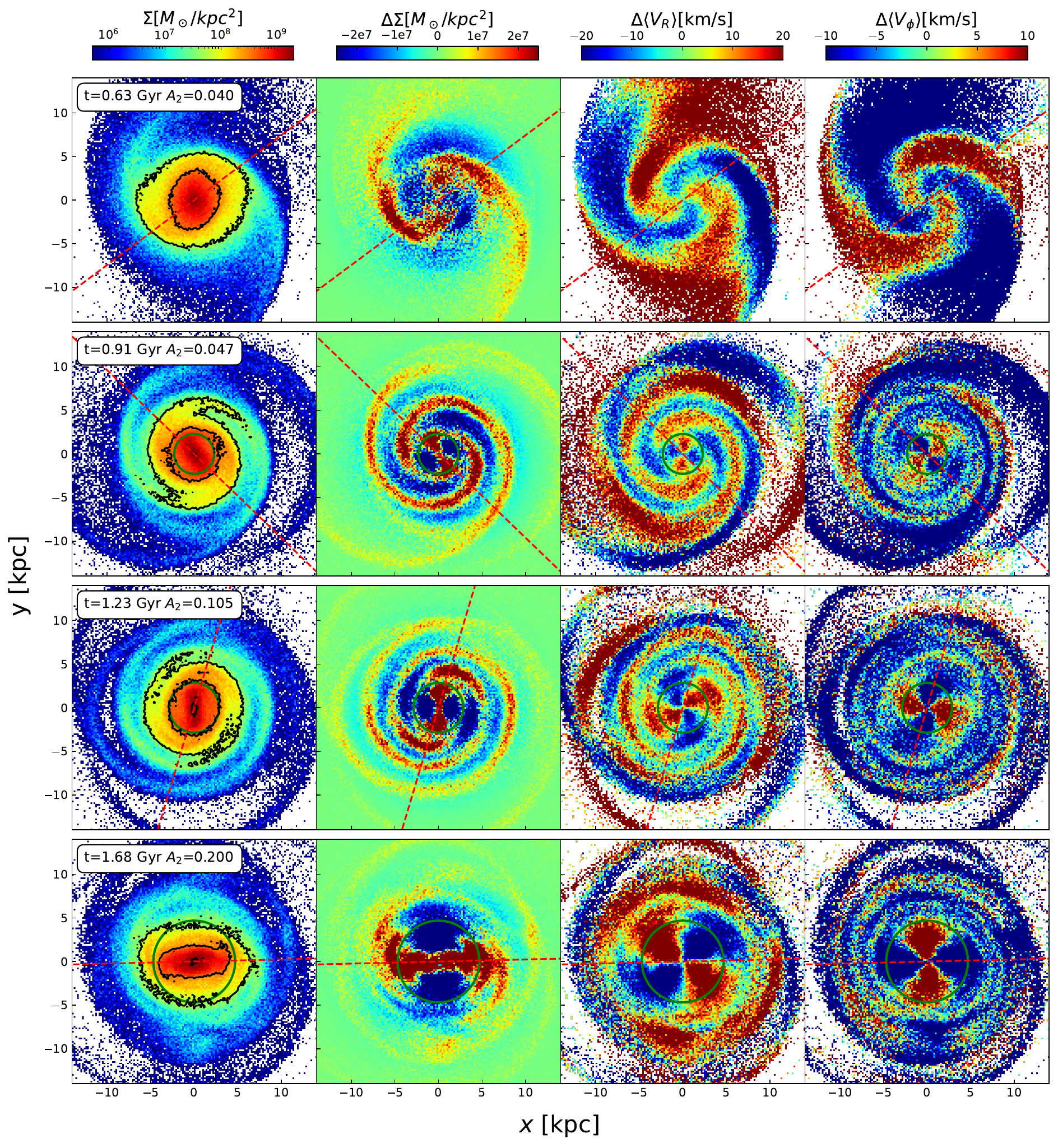}
  \caption{
    The onset of a large-scale bar-like perturbation during a flyby interaction (warm disk, 1/3, perpendicular orbit). 
    The left column displays the stellar surface density maps with the time and $A_2$ amplitude indicated in the text box. 
    The right three columns show the differences between the current snapshot and the initial condition in terms of stellar surface density ($\Delta\Sigma$), averaged radial velocity ($\Delta\langle V_R \rangle$), and averaged azimuthal velocity ($\Delta\langle V_\phi \rangle$), respectively. 
    The rows present a temporal evolution from top to bottom. 
    In the first row, the tidal perturbation wave propagates to the galaxy center shortly after the perturber's pericenter passage. This wave initiates a ``seed bar'' as shown by the quadrupole signal in the velocity fields (outlined by green circles) in the second row. The third and fourth rows illustrate the bar's evolution at times when $A_2$ reaches 0.1 and 0.2, respectively. In each panel, the $\phi$ angle of the $A_2$ signal is marked with a red dashed line.
    }
  \label{fig:per_wave}
\end{figure*}

Although the perturbation does not alter the bar growth rate, it plays a crucial role in triggering bar formation. 
In most of our cases, the perturber promotes the onset of bar formation. 
Following the works of \citet{Dubinski2008}, \citet{Polyachenko2016}, and \citet{Moetazedian2017}, we investigate how interactions influence the onset of bar formation by examining the perturbation waves.

% We analyze the same flyby interaction shown in \autoref{fig:x_evo} (warm disk, \massratioeq{1/3}, perpendicular orbit), in which the bar formation is advanced by 0.7\Gyr.
We analyze the flyby interaction of the warm disk model under an intermediate perturbation (\massratioeq{1/3}, perpendicular orbit), in which the bar formation is advanced by 0.7\Gyr.
The onset of the large-scale bar-like perturbation in this case is presented in \autoref{fig:per_wave}. 
The left column shows stellar surface density maps.
The right three columns illustrate the deviations between the current snapshot and the initial condition, specifically in terms of stellar surface density ($\Delta\Sigma$), averaged radial velocity ($\Delta\langle V_R \rangle$), and averaged azimuthal velocity ($\Delta\langle V_\phi \rangle$), respectively. 
We note that the average radial velocity of the axisymmetric disk at the beginning of the simulation is zero.
% theoretically zero by Jeans theorems \citep[see e.g.][section 4.2]{Binney2008}.
% Thus, the third column can also be interpreted as the radial velocity field $\langle V_R \rangle$.}
Thus, $\Delta\langle V_R \rangle = \langle V_R \rangle$.

The first row depicts a time shortly after the perturber's closest approach. 
The interaction triggers a perturbation wave that propagates to the galaxy's center. 
This wave generates a ``seed bar'', as indicated by the quadrupole signal in the velocity fields shown in the second row. 
Subsequently, the swing amplification feedback loop amplifies this seed bar into a robust bar structure. 
We illustrate the bar's evolution at $t_0$, i.e., the time when $A_2$ reaches 0.1, in the third row. 
The final row displays the bar at a later stage when $A_2$ reaches 0.2, a common criterion for identifying strong bars.

To further confirm the link between the perturbation and the onset of the bar formation, we rerun several flyby simulations of the warm disk with the pericenter time being delayed by 1\Gyr. 
As expected, the bar formation is delayed by a similar amount compared with the original runs, as revealed by the $t_0$ parameter.

\begin{figure*}
  \centering
  \includegraphics[width=0.95\textwidth]{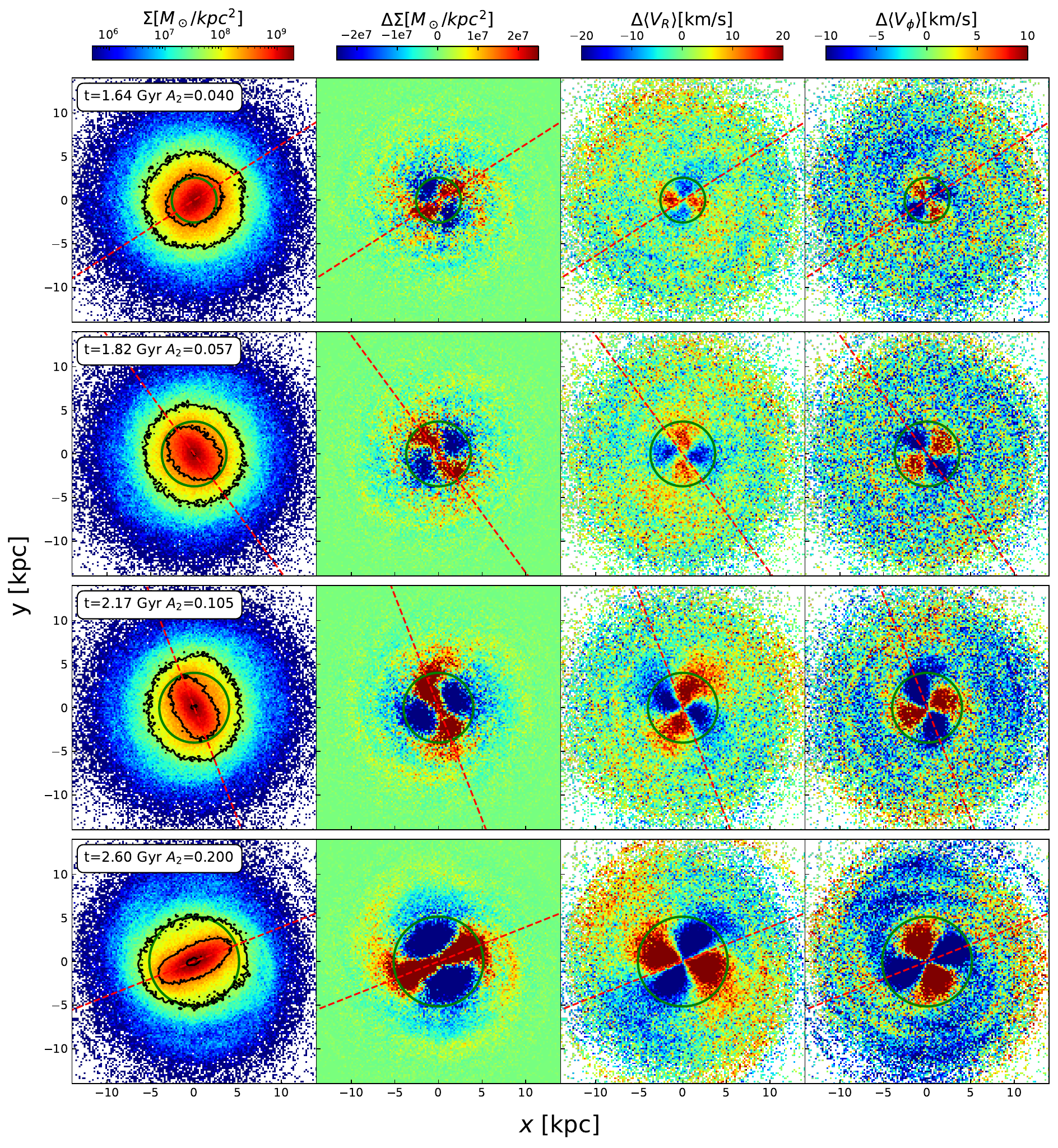}
  \caption{Same as \autoref{fig:per_wave} but for the same disk evolved in isolation. The ``seed bar'' spontaneously emerges at the galaxy center and gradually evolves into a strong bar structure. The ``seed bar'' emerges $\sim0.7$\Gyr\ later than the flyby interaction case shown in \autoref{fig:per_wave}.}
  \label{fig:iso_wave}
\end{figure*}

For comparison, we present the onset of bar formation in the same disk evolved in isolation in \autoref{fig:iso_wave}. 
A similar quadrupole signal in the velocity fields spontaneously emerges at the galaxy center as shown in the first row.
However, the ``seed bar'' appears 0.7\Gyr\ later compared to the flyby interaction scenario shown in \autoref{fig:per_wave}. 
The bottom three rows of \autoref{fig:iso_wave} illustrate the development of the ``seed bar'' into a strong bar structure. 
The evolution of the bar in the isolated case is similar to that under tidal perturbation (\autoref{fig:per_wave}) except for the stronger spiral arms in the latter case.
This consistent evolution further supports the notion that the tidal bar grows at a rate comparable to its spontaneous counterpart in the warm disk.

\begin{figure*}
  \centering
  \includegraphics[width=0.95\textwidth]{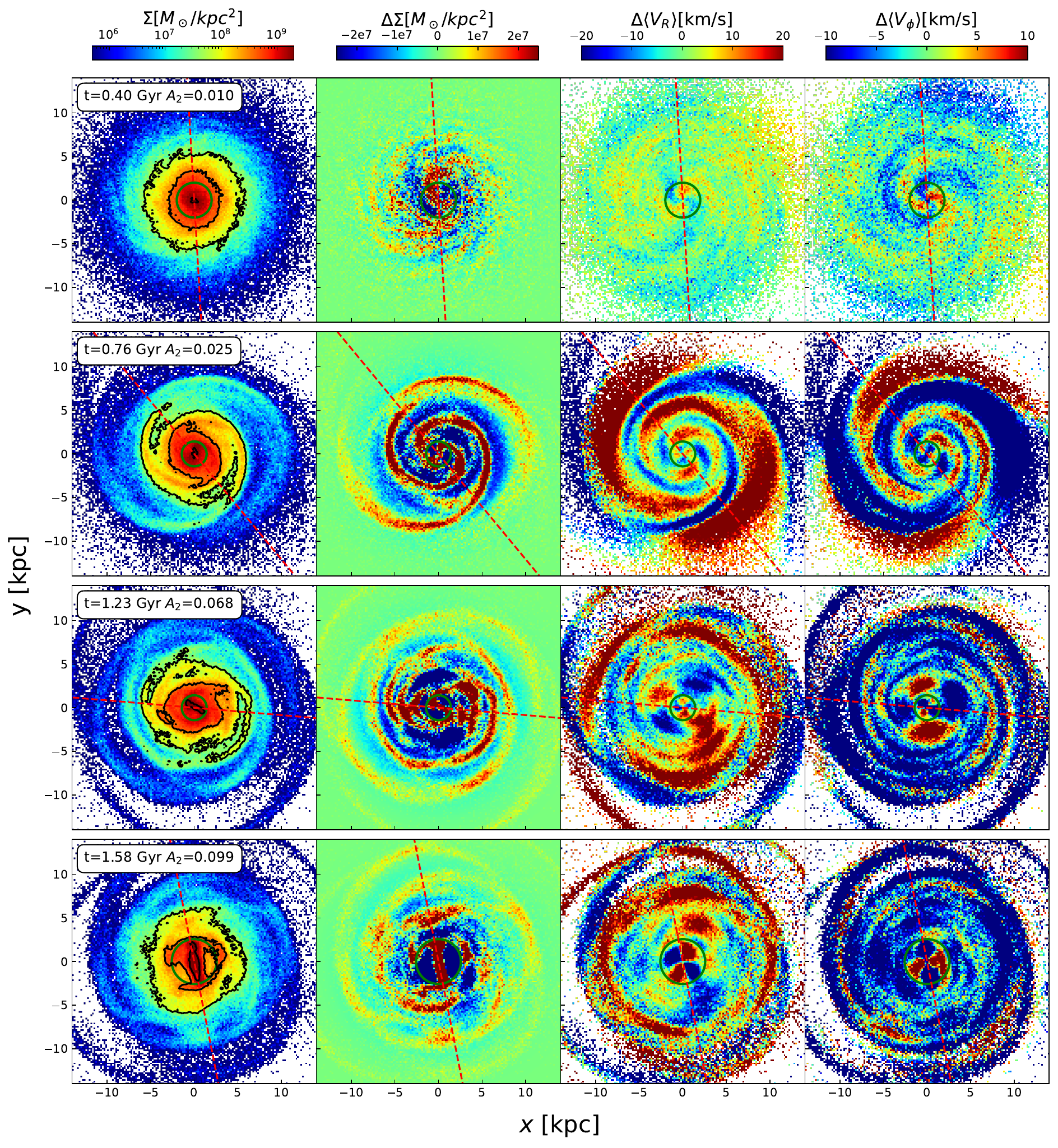}
  \caption{Same as \autoref{fig:per_wave} but for a delayed case (cold disk, 1/3, perpendicular). 
  Although weak, the cold disk spontaneously generates a ``seed bar'' prior to the interaction (top row). When the perturbation wave reaches the galaxy's center, it is out of phase with the preexisting seed bar, causing destructive interference and limiting the bar's growth (middle two rows). Despite this initial delay, the bar eventually resumes steady growth at a later stage (bottom row).
  }
  \label{fig:delay_wave}
\end{figure*}

Tidal perturbations can also postpone the onset of bar formation. 
\autoref{fig:delay_wave} illustrates a delayed scenario in the cold disk model (\massratioeq{1/3}, perpendicular orbit). 
The first row shows a moment shortly before the perturber's closest approach. 
Although weak, the cold disk has already spontaneously generated a ``seed bar'' prior to the interaction. 
When the perturbation wave reaches the galaxy center, it interferes destructively with the original seed bar, which twists the quadrupole signal in the velocity fields of the central region and
disrupts the spiral arms in the outskirts. The destructive interference limits the bar's growth. 
Ultimately, one mode prevails over the other at a later time and starts the steady growth of the bar as seen in the bottom row ($A_2=0.1$). 
We propose that the internal mode dominates over the external mode as the perturber is already more than $300\; R_d$ away when \BS\ reaches 0.1.

% We align with the interpretation of \citet{Moetazedian2017} for the delayed case.
% The perturbation wave is not in phase with the internally developed seed bar when it arrives at the galaxy center, leading to destructive interference.
% We test this idea by rerunning the delayed case with a slightly different pericenter time by $\pm0.1$\Gyr. 
% Both the earlier and later interactions promote the onset of bar formation by about 0.2\Gyr\ compared with the isolated cases instead of delaying it in the original interaction.

Our results align with the interpretation of \citet{Moetazedian2017} regarding the delayed case. 
Destructive interference is due to the external perturbation wave being \textbf{accidentally} out of phase with the internally developed seed bar when it reaches the galaxy center.
We test this hypothesis by resimulating the delayed case with slightly altered pericenter times, shifted by $\pm0.1$\Gyr. In both the earlier and later interactions, the onset of bar formation is advanced by approximately 0.2\Gyr\ compared to the isolated cases, rather than being delayed as in the original interaction.
This result underscores the difficulty for tidal perturbations to delay bar formation within galaxies inherently susceptible to internal instability, thereby explaining why the majority of our interaction simulations have earlier bar emergences. Tidal perturbations provide stronger initial noises for the swing amplification feedback loop, while the probability of the external perturbation wave being out of phase with the internally developed seed bar is indeed small.
Our results also highlight the complexity of predicting the galaxy morphology, which is influenced by both their environment and internal baryonic physics \citep{Zhou2020}. 

% \revision{Tidal perturbations have the smaller probability to delay the bar formation than to advance it if the galaxy is prone to bar instability.}

\subsection{bar growth in the hot disk model}
\label{sec:hot_disk}

% In a more specific context, the term ``tidally \revision{forced} bars'' refers exclusively to bars in galaxies that are stable against bar formation when isolated.
As for bars in galaxies that are stable against bar formation when isolated (``tidally forced bars''), it is unclear whether their growth timescales are similar regardless of the perturbation strength. 
Consequently, we examined bar growth in the hot disk model, which avoids bar instability throughout a 6\Gyr\ isolated evolution.

\begin{figure*}
  \centering
  \includegraphics[width=\textwidth]{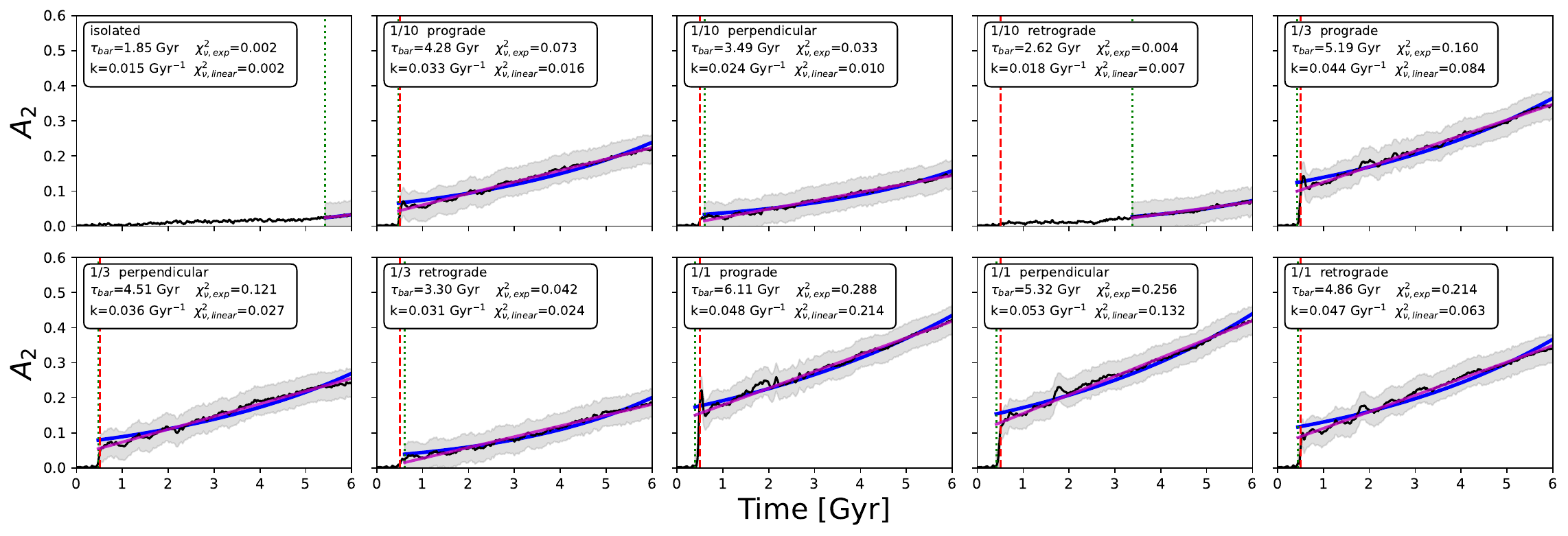}
  % \caption{test}
  \caption{Similar to \autoref{fig:exp_fit} but for the hot disk model. The magenta lines show the linear fitting results. The text box provides the simulation information alongside the exponential growth timescale \tbar\ and the linear growth rate $k$. $\chi_\nu$ of different fittings are also noted. Unlike the cold and warm disks, the hot disk does not show consistent \tbar\ values. 
  }
\label{fig:hot_fit}
\end{figure*}

We utilize the {\bf same exponential fitting (\autoref{eq:exp_fit})} for the $A_2$ evolution in the hot disk model. 
Since $A_2$ does not exhibit a distinct peak in the hot disk, we extend the upper limit of the fitting time range to the end of the simulation, i.e., 6\Gyr. 
For the lower limit, we set $A_2>0.025$ and exclude the initial 0.75\Gyr\ of the simulation. 
Visual examination of the stellar surface density indicates that the significant increase in the $A_2$ signal around the interaction time ($t_{\rm peri}=0.5$\Gyr) is primarily due to overall distortion of the disk and transient spirals, rather than bar formation. 
The fitting outcomes are presented in \autoref{fig:hot_fit}. 
Unlike the cold and warm disks, the hot disk does not display a consistent \tbar, ranging from a minimum of 1.85\Gyr\ to a maximum of 6.11\Gyr. Notably, the \massratioeq{1/1} interactions yield larger \tbar values compared to the \revision{interactions with less massive perturbers}, even though the bars in the former cases are stronger.
\revision{When \massratio\ is fixed, the \tbar\ values are larger for bars developed under prograde interactions compared to those that form under perpendicular and retrograde interactions.
Since perturbers on prograde orbit provide stronger perturbations compared to those in perpendicular or retrograde orbit (see \citeauthor{Lokas2018} \citeyear{Lokas2018} and Sec~4.2 of \citetalias{Zheng2025}), this result also implies that stronger perturbations result in larger \tbar\ values.
We suggest that the \tbar\ of the tidally forced bar qualitatively correlates with the strength of the perturbation, and leave quantification of the correlation for further investigation in the future.}
\footnote{\revision{One way to measure the perturbation strength by the dimensionless tidal strength parameter $S$ introduced in \cite{Elmegreen1991}.
We found that $S$ is approximately 0.87, 0.15, and 0.03  for \massratioeq{1/1}, 1/3, and 1/10 respectively. However, $S$ does not take into account the inclination of the perturber's orbit that also affects the strength of tidal interaction (see \citeauthor{Lokas2018} \citeyear{Lokas2018}; Sec~4.2 of \citetalias{Zheng2025}).}}

The growth of $A_2$ in the hot disk does not adhere to a clear exponential pattern, unlike in the cold and warm disks. 
% A linear function appears more suitable for describing their growth. We employ the following function to fit the $A_2$ growth:
In addition to exponential fitting, we also utilize the following linear fitting to model the evolution of $A_2$ in the hot disk model:
\begin{equation}
A_2 (t) = k t + b,
\end{equation}
where $k$ represents the linear growth rate of the bar strength. 
The linear fitting is also presented in \autoref{fig:hot_fit} with the $k$ values noted in the text box. 
\revision{Similar to the result for \tbar, we find a qualitative relation} that stronger interactions correspond to larger $k$ values, suggesting a correlation between interaction strength and bar growth in the hot disk.

Visually, the linear function seems to offer a slightly better fit to \revision{the growth of tidally forced bars} in the hot disk for most scenarios. 
From a statistical perspective, the linear fit yields a smaller reduced chi-squared statistic ($\chi_\nu^2$) compared to the exponential fit. 
Conversely, the cold and warm disks exhibit a more favorable fit with the exponential function during the bar formation phase. The linear fitting approach may be more appropriate for describing the secular evolution of the bar in the cold and warm disks, occurring after the buckling stage and prior to the saturation stage (refer to the inset axes in \autoref{fig:exp_fit}, and also Figure 4 in \citetalias{Zheng2025}).

\revision{The linear growth can be interpreted as an approximation of exponential growth with a large timescale. 
If a series of models of increasing stability is simulated, it is expected that the increasing \tbar\ will gradually transition the bar growth from exponential to linear. 
Therefore, the slight preference for linear bar growth alone is insufficient to serve as decisive evidence that tidally forced bars in the hot disk model are fundamentally different from those developed in the cold and warm disks.}

% The tidally \revision{forced} bars in the hot disk deviate from the exponential growth profile.
% , with growth rate \revision{(both the exponential and linear ones)} being linked to the interaction strength. 
% This contrasts 
\revision{However, we emphasize an important fact that both the exponential and linear growth rates of the tidally forced bars in the hot disk correlate with the perturbation strength. This contrasts
with the consistent growth rate of bars within cold or warm disks}, where bar growth is primarily driven by internal disk instability. 
This difference makes it less meaningful to compare the growth timescale \tbar\ between the hot disk and the cold/warm disks. 
It also suggests that bars in the hot disk may not operate under the \revision{swing amplification mechanism, contrasting with those in the cold and warm disks}. 
Possibly, the hot dynamical state of the disk resists the disturbance caused by the perturber, preventing disruption of the galaxy center and rendering it less susceptible to the swing amplification feedback loop \citep{Dubinski2008}.

% \textcolor{red}{We compute the swing amplification parameter $X$ for the hot disk model. Following the interaction, the $X$ parameter exhibits a slow yet steady increase, which could be attributed to the distortion of the disk. }
We examine the velocity fields of the bars within the hot disk model and observe that the quadrupole signal is more extended but weaker compared to bars in the cold and warm disks with similar bar strength. 
Bars in hot disks are also longer and less eccentric at the same $A_2$. Their subsequent evolution encompasses both the elongation of the bar and the increase in eccentricity.

% If the term ``tidally \revision{forced} bars'' is strictly applied to those in galaxies that inherently avoid bar instability, their formation and evolution differ from those in galaxies that are prone to bar instability.
The formation and evolution of  tidally forced bars differ from those in galaxies that are prone to bar instability. 
Quantitative comparison of bar shapes and dynamical/kinematical analysis of the bar regions have the potential to distinguish these two categories of bars. This will constitute the long-term objective of this series of papers.

%-----------
%-- Sect. 5
%-----------

\section{Summary}
\label{sec:summary}

We employ a suite of controlled $N$-body simulations to investigate the growth of bars under both tidally \revision{induced} and internally developed mechanisms.
We generate three pure disk galaxy models with varying stabilities against bar formation by adjusting the radial velocity dispersion.
The models are labeled cold, warm, and hot. 
The cold and warm disk models spontaneously form bars, although the warm one takes a longer time to do so. The hot disk model does not develop a bar within 6\Gyr\ of isolated evolution.
We then introduce a perturber to induce tidal interactions with mass ratios of 1/1, 1/3, and 1/10 and inclination angles of 0\degree, 90\degree, and 180\degree\ relative to the disk plane.

We apply an exponential fitting to the $A_2$ evolution in cold and warm disks to quantify the growth timescale \tbar\ and the onset time $t_0$ of the bar formation (\autoref{fig:exp_fit}).
The tidally \revision{induced} bars show similar \tbar\ to that of the internally developed bars with differences less than 20\%, indicating that tidal bars have similar growth rates to their spontaneous counterpart in the same disk (\autoref{fig:fit_results}).
The tidal perturbation promotes(delays) the bar formation by advancing(postponing) the onset time $t_0$.
This conclusion is further corroborated by comparing the $A_2$ evolution against \revision{$(t-t_{0})$. } (\autoref{fig:shifted_a2}).
If the disk inherently possesses instability, the primary driver of bar formation is its internal nature, which remains largely unaffected by tidal perturbations.
%  \textcolor{red}{ as indicated by the swing amplification parameter $X$ (\autoref{fig:x_evo}).}

We further investigate the onset of the bar formation by examining the perturbation wave.
In instances where the perturbation advances bar formation, the perturbation wave propagates to the galaxy center shortly after the perturber's pericenter and evokes a ``seed bar'' (\autoref{fig:per_wave}).
The isolated case shows a similar evolution of the ``seed bar'' but its spontaneous emergence occurs approximately 0.7\Gyr\ later than in the interaction case (\autoref{fig:iso_wave}).
We also show a delayed interaction case where the perturbation wave is out of phase with the internally developed seed bar, leading to destructive interference and limiting the growth of the bar (\autoref{fig:delay_wave}).

In the hot disk model, the tidally \revision{forced} bars do not show consistent \tbar\ values (\autoref{fig:hot_fit}), probably due to deviations from exponential growth profiles.
Linear fitting provides a slightly better description of the bar growth in the hot disk than the exponential fitting.
\revision{These results make} it less meaningful to compare the growth timescale \tbar\ between the hot disk and the cold/warm disks and hints that bars in the hot disk may not follow the same growth mechanism as those in the cold and warm disks.

% However, the difference in the bar growth mechanisms holds the potential to separate the strictly-speaking ``tidally \revision{forced} bars'' from those developed in galaxies that are inherently susceptible to bar instability.
However, the difference in the bar growth mechanisms holds the potential to separate \revision{tidally forced bars that developed in stable galaxies} from those developed in galaxies that are inherently susceptible to bar instability.
Further investigation of the bar's shape and dynamical/kinematical analysis of the bar regions could be important in distinguishing between these two types of bars.

\revision{
With the advent of powerful telescopes, such as the Atacama Large Millimeter Array and the James Webb Space Telescope, astronomers can now observe the gas-rich phase of galaxies at high redshifts.
Considering that all disk galaxies undergo a gas-rich phase, the new era of gas-rich controlled simulations is vital for further insights into the bar formation mechanism \citep{BlandHawthorn2024,BlandHawthorn2025}. We will explore the impact of gas on bar formation and evolution under both the internal and external formation mechanisms in future work.}

\software{
    {\sc agama}\citep{AGAMA2019},
    \texttt{GADGET-4} \citep{Springel2005,Springel2021}. 
    NumPy \citep{2020NumPy-Array},
    SciPy \citep{2020SciPy-NMeth},
    Matplotlib \citep{4160265},
    Jupyter Notebook \citep{Kluyver2016jupyter}
}

% \begin{acknowledgments}
\section*{Acknowledgements}
We thank the referee for suggestions that helped to improve the presentation of the paper. 
\revision{We thank Thor Tepper-Garcia and Joss Bland-Hawthorn for their help with the \agama\ work and with initializing the galaxy models.}
We thank Zhi Li and Sandeep Kumar Kataria for their valuable insights on simulations and analysis. We also thank Rui Guo for their helpful discussions.
% FUNDING!!!
The research presented here is partially supported by the National Key R\&D Program of China under grant No. 2018YFA0404501; by the National Natural Science Foundation of China under grant Nos.  12025302, 11773052, and 11761131016; by the ``111'' Project of the Ministry of Education of China under grant No. B20019; and by China Manned Space Program with grant No. CMS-CSST-2025-A11. J.S. acknowledges support from the {\it Newton Advanced Fellowship} awarded by the Royal Society and the Newton Fund. 
X.W. wishes to thank the financial support from the Natural Science Foundation of China (Numbers NSFC-12073026 and NSFC-12433002).
This work made use of the Gravity Supercomputer at the Department of Astronomy, Shanghai Jiao Tong University.
% \end{acknowledgments}

%%%%%%%%%%%%%%%%%%%%%%%%%%%%%%%%%%%%%%%%%%%%%%%%%%

%%%%%%%%%%%%%%%%% APPENDICES %%%%%%%%%%%%%%%%%%%%%

\appendix
\restartappendixnumbering
\section{Fitting validation}
\label{app:diff_fit}

% different fitting formulas yield similar $\tau_{\mathrm{bar}}$--> appendix, figure
In addition to \autoref{eq:exp_fit}, we explored alternative exponential formulas to fit the $A_2$ evolution. One such formula is
\begin{equation}
A_2 (t) = C_1 \exp(t/\tau_{\mathrm{bar}}) + C_2,
\end{equation}
which offers greater flexibility in the fitting process. 
We also examined the following formula
\begin{equation}
A_2 (t) = C (\exp(t/\tau_{\mathrm{bar}}) -1 ),
\end{equation}
This choice is driven by the consideration that the bar strength \revision{$A_2\approx 0$ at $t=0$ if the disk is set to be perfectly axisymmetric at the beginning of the simulations,}
\revision{even though \BS\ is not expected to stay at zero for a long time. }

\begin{figure}
  \centering
	\includegraphics[width=0.6\columnwidth]{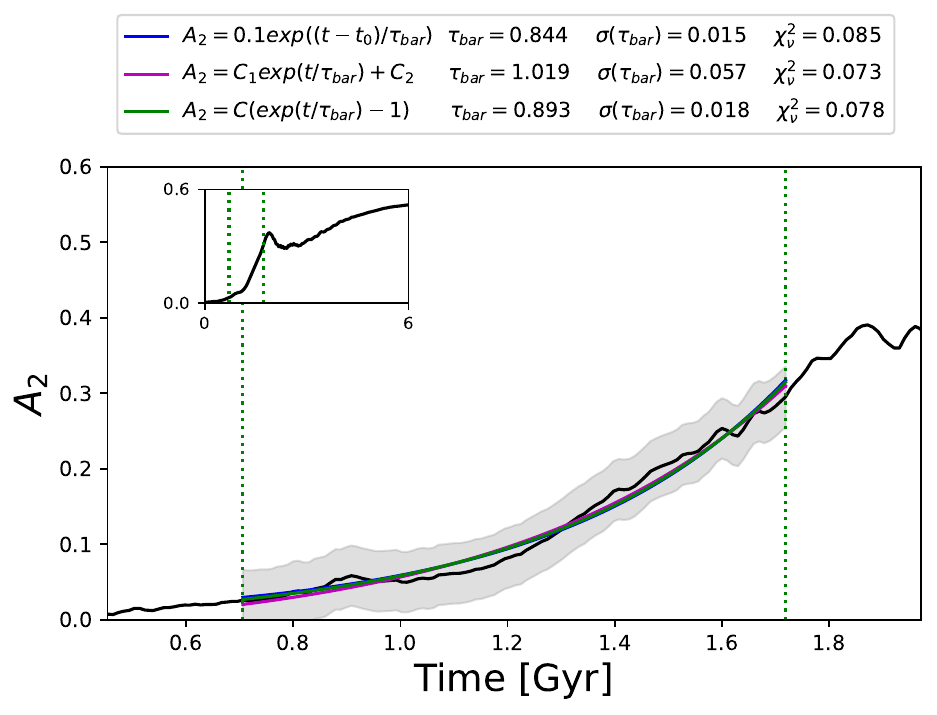}
	\caption{
    Results of exponential fitting using different formulas for the isolated evolution of the cold disk. The legend notes the formulas used, the growth timescale \tbar, the $1\sigma$ error, and the reduced chi-squared statistic $\chi_\nu^2$.  Different formulas yield similar \tbar\ with a difference ~20\% while having the same ``goodness'' of fitting as shown by the fitted lines and the similar values of $\chi_\nu^2$.}
	\label{fig:diff_fit_fig}
\end{figure}

We employ the same fitting procedure outlined in Section \ref{sec:exp_growth} with these alternative formulas to fit the $A_2$ evolution of the internally developed bar in the cold disk model. 
The results are presented in \autoref{fig:diff_fit_fig}. 
Despite employing different formulas, the fits exhibit a similar ``goodness'', as indicated by the fitted lines and the comparable $\chi_\nu^2$. 
The growth timescales \tbar\ derived from these formulas are also similar, with differences of $\sim$20\%. 
Consequently, we consider a difference of $\sim$20\% in \tbar\ to be acceptable.

\begin{figure}
  \centering
  \includegraphics[width=0.6\columnwidth]{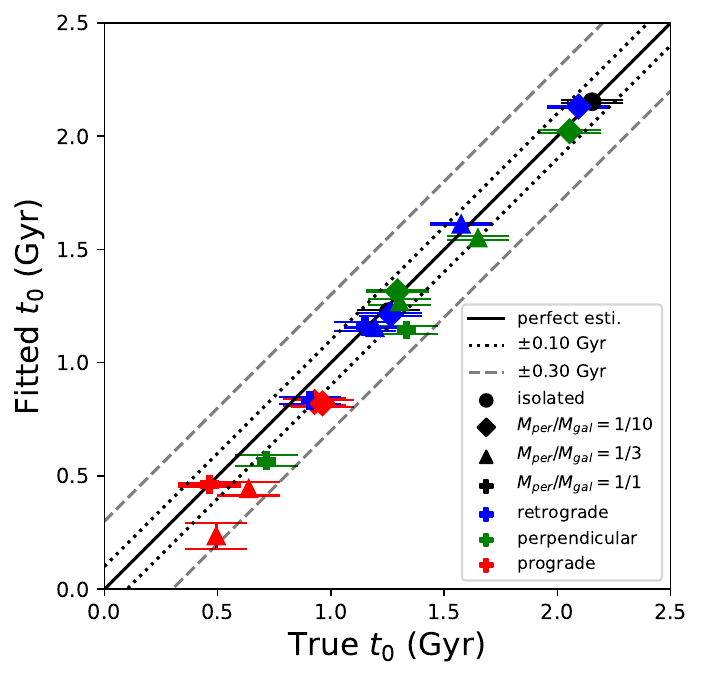}
  % \caption{test}
  \caption{Comparison of the fitted $t_0$ and true $t_0$. Each point represents a simulation, with the shape indicating the mass ratio and the type representing the orbit configuration. The solid diagonal line is the perfect estimation of  $t_{0, \mathrm{fit}}=t_{0, \mathrm{true}}$. The dotted and dashed lines mark the offsets of $\pm0.1$\Gyr\ and $\pm0.3$\Gyr, respectively.
  The reasonable agreement between fitted $t_0$ and true $t_0$ outlines the reliability of the exponential fitting.
  % Most simulations show nice agreement within $\pm0.1$\Gyr\ with the other 6 simulations having slightly larger offset that is between 0.1 and 0.3 \Gyr. 
  }
  \label{fig:check_t0}
\end{figure}

To further demonstrate the reliability of the exponential fitting, we compare the fitted $t_0$ values with the true $t_0$ values in \autoref{fig:check_t0}. 
\revision{The fitted $t_0$ is defined as the time at which the fitted $A_2$ profile reaches 0.1, while the true $t_0$ is the time when the actual $A_2$ profile attains this value. }
The majority of the simulations show good agreement within $\pm0.1$\Gyr\ between and fitted and true $t_0$.
Only six simulations show slightly larger offsets, ranging between 0.1 and 0.3 \Gyr. 
This agreement underscores the robustness of the exponential fitting approach.

% \newpage

\bibliography{tidal_bar.bib}
\bibliographystyle{aasjournal}

\end{document}